# Prompt and afterglow early X-ray phases in the commoving frame. Evidence for Universal properties?

*G. Chincarini[1,2], A. Moretti[2], P. Romano[2], S. Covino[2], G. Tagliaferri[2], S. Campana[2], M. R. Goad[3], S. Kobayashi[4], B. Zhang[5], L. Angelini[6], P. Banat[2], S. Barthelmy[6], A. Beardmore.[3], P.T., Boyd[6], A. Breeveld[7], D. N. Burrows[4], M. Capalbi[8], M.M. Chester[4], G. Cusumano[9], E. E. Fenimore[10], N. Gehrels[6], P. Giommi[8], J. E. Hill[6], D. Hinshaw[6], S.T. Holland[6], J. A. Kennea[4], H.A. Krimm[6], V. La Parola[9], V. Mangano[9], F.E. Marshall[6], K.O. Mason[7], J. A. Nousek[4], P. T. O'Brien[3], J. P. Osborne[3], M. Perri[8], P. Mészáros[4], P.W.A. Roming[4], T. Sakamoto[6], P. Schady[6], M. Still[6], A.A. Wells[3].*

[1] *Università degli studi di Milano-Bicocca, Dipartimento di Fisica, Italy*
[2] *INAF -Osservatorio Astronomico di Brera, Via E. Bianchi 46, 23807 Merate (LC), Italy*
[3] *Department of Physics and Astronomy, University of Leicester, Leicester, LE 17 RH,UK*
[4] *Department of Astronomy & Astrophysics, Pennsylvania State University, USA*
[5] *Department of Physics, University of Nevada, Las Vegas, NV 89154-4002, USA*
[6] *NASA - Goddard Space Flight Centre, USA*
[7] *Mullard Space Science Laboratory, University College London, UK*
[8] *ASI Science Data Center, via G. Galilei, 00044, Frascati, Italy*
[9] *INAF- Istituto di Astrofisica Spaziale e Fisica Cosmica Sezione di Palermo, Italy*
[10] *Los Alamos National Laboratory, New Mexico, USA*
[11] *Universities Space Research Association*

## Abstract

We analyze the Swift XRT light curves and spectra of the gamma-ray bursts (GRBs) for which the redshift has been measured. The sample consists of seven GRBs. The soft X-ray light curves of all these GRBs are separated into two morphological types 1) those starting off with a very steep light curve decay and 2) those showing a rather mild initial decay. In the first case the initial decay is followed by a flattening and by a further steeping while in the second case the mild decay is followed by a steeper decay. During these transitions the soft X-ray spectrum of these GRBs remains constant within the observational errors (except for one case, GRB050319). For the first time we are able to exploit the early light curve of GRB afterglows in the co-moving frame. Aside from their temporal and spectral behavior, we find that the energy of the afterglow emitted in the (rest frame) time interval 20-200 s and 1300-12600 s after the BAT trigger, correlates with the mean energy of the prompt emission, hinting for a close link between the two. The flux emitted in the soft X-rays (0.2-10 keV), soon after the burst, goes from 6% to about 20% of the prompt emission while the XRT rest frame energy emitted in the soft X ray is 16% to 42% of the hard X-ray flux (15 - 350 keV) observed during the prompt emission phase.

## 1. Introduction.

Following the discovery of GRB afterglows by BeppoSAX (Costa et al. 1997) and the subsequent development of theoretical models to explain these spectacular cosmic explosions (e.g. Mészáros and Rees 1997 and references therein), a clear picture of the

GRB phenomenon has emerged. For a thorough review of the excellent early work in this field see the review papers by Piran (2004), Hurley, Sari and Djorgovski (2003), Zhang and Mészáros (2004).

The primary goal of the Swift Gamma-ray Burst Explorer (Gehrels et al. 2004) launched Nov 20, 2004 is not only to detect a statistically significant sample of GRBs, but also to collect related data in the X-ray (0.2-10 keV) and optical (1700 – 6500 Å) bands in the as-yet little understood initial few tens of seconds after the trigger.

At the present time, Swift is collecting a large amount of new data, with seemingly every burst displaying its own peculiar characteristics. Indeed, with every new prompt XRT observation of a burst, new details are emerging (e.g. Burrows et al. 2005a; Campana et al. 2005a; Tagliaferri et al. 2005; Goad et al. 2005; Cusumano et al. 2005; Burrows et al. 2005b; Vaughan et al. 2005).

To determine whether we can classify GRBs based on the characteristics of their light curves, and thereby obtain an understanding of the physical mechanism behind the observed differences, we report here the results of a pilot study of a small, but statistically significant sample of GRBs for which we have both early observations in the XRT and a spectroscopic redshift for the optical transient. By choosing the sample in this fashion we can attempt to disentangle intrinsic physical properties from those related to GRB distance. This study will serve as the basis for the analysis of a much larger sample of GRBs for which more significant statistics will be possible.

The paper is organized as follows: in section 2 we discuss the observations while in section 3 we detail the analysis. Finally in section 4 we discuss our results. Throughout this paper the decay and the spectral indices are parametrized as follows: $F_\nu \propto t^{-\alpha} \nu^{-\beta}$ where t is time and ν is the frequency.

## 2. The observations

The present sample includes all GRBs (seven) observed by Swift from January 2004 to May 15 for which spectroscopic redshifts of their host galaxies are available. Of these seven GRBs, six were discovered by the Burst Alert Telescope (BAT, Barthelmy et al. 2004) on board Swift and the other by HETE2. These seven objects are listed in Table 1.

Here we concentrate on X-ray observations carried out with the X-Ray Telescope (XRT) on board Swift (Burrows et al. 2005c). The XRT CCD has been designed to automatically switch between observing modes so that bright sources may be observed without any pile-up effect (this occurs when two photons hit the same pixel during a single CCD frame). For a thorough description of XRT observing modes and switch points see Hill et al. (2004). However, in the early stages of the mission some of the GRB afterglows interrupted ongoing XRT calibration observations of known sources. For these observations XRT was in manual state and hence on slewing to the burst position the normal sequence of observations was not performed. For the few GRBs affected in this manner, the X-ray observations were usually performed in photon counting (PC) mode. Consequently, for the brightest sources the XRT data are severely piled-up. The effects of pile-up can be corrected for by extracting light curves and spectra from an anular region around the source center (rather than a simple circular region), with a 'hole', the size of which is a function of the source brightness (see e.g. Vaughan et al. 2005, for a thorough

Table 1

| Burst | Redshift | T90 (*) | Fluence 15 – 350 keV erg cm$^{-2}$ | References |
|---|---|---|---|---|
| GRB050126 | 1.29 | 25.8 | $2.0 \times 10^{-6}$ | 1 |
| GRB050315 | 1.95 | 96 | $3.1 \times 10^{-6}$ | 2 |
| GRB050318 | 1.44 | 31 | $1.5 \times 10^{-6}$ | 3, 2 |
| GRB050319 | 3.24 | 160.5 (&) | $1.7 \times 10^{-6}$ | 4, 2 |
| GRB050401 | 2.9 | 34.0 | $1.5 \times 10^{-5}$ | 5 |
| GRB050408 | 1.24 | / | / | 6 |
| GRB050505 | 4.27 | 62.0 | $4.4 \times 10^{-6}$ | 7 |

&) The BAT light curve shows two flares. The second with $T_{90}$ = 23.5 s, fluence $7.3 \times 10^{-7}$ erg cm$^{-2}$ and the first preceding it by 137 seconds, fluence $1.6 \times 10^{-6}$ erg cm$^{-2}$.

(*) T-90 is the time needed to accumulate from 5% to 95% of the counts in the 15-350 keV band.

1) Campana et al (2005b), Sato et al. 2005, Berger et al. 2005a ; 2) Beardmore A.P., et al. 2005, Parsons., et al. 2005, Kelson,, et al. 2005; 3) Krimm,, et al. 2005a, Berger,, et al. 2005b; 4) Krimm, et al. 2005 b, Fynbo, et al., 2005a ; 5) Angelini, et al 2005, Markwardt, et al 2005, Fynbo, et al 2005b ; 6) Chincarini, et al 2005, Sakamoto, et al 2005, Prochaska, et al 2005 ; 7) Hullinger, et al., 2005.

discussion of how to mitigate the effects of pile-up in the worst affected source GRB050315). The total flux is then recovered by using the PSF analytical model. discussion of how to mitigate the effects of pile-up in the worst affected source GRB050315). The total flux is then recovered by using the PSF analytical model.As the afterglow decays the pile-up effect diminishes and extraction from a circular region is then feasible. For the last four bursts discussed here, automatic mode switching was enabled, and for three of them, observations started in windowed-timing (WT) mode (providing just 1D imaging) due to high initial count rates. Cross-calibration between modes ensures that the two modes (ie. PC and WT) do not introduce artificial variations in flux. In our analysis we extracted the X-ray spectra and light curves from a circular region of 30 pixels radius for PC mode, and a rectangular region of 40x20 pixels in WT mode, centered on the source position as determined from the Swift analysis task **xrtcentroid**. The extraction regions for the cases where pile-up is significant are detailed below where we discuss individual sources. The XRT background (which is in any case very low due to the low Earth orbit of Swift) must be taken into consideration especially at low count rates . We evaluate the background over an annulus centered on the GRB position and delimited by 50 and 90 pixels (internal and external radius, respectively), taking care to avoid known background sources. In WT mode we used a 40x20 pixel region at a distance of 50 pixels from the source. The mode in which the various sources were observed is described below:

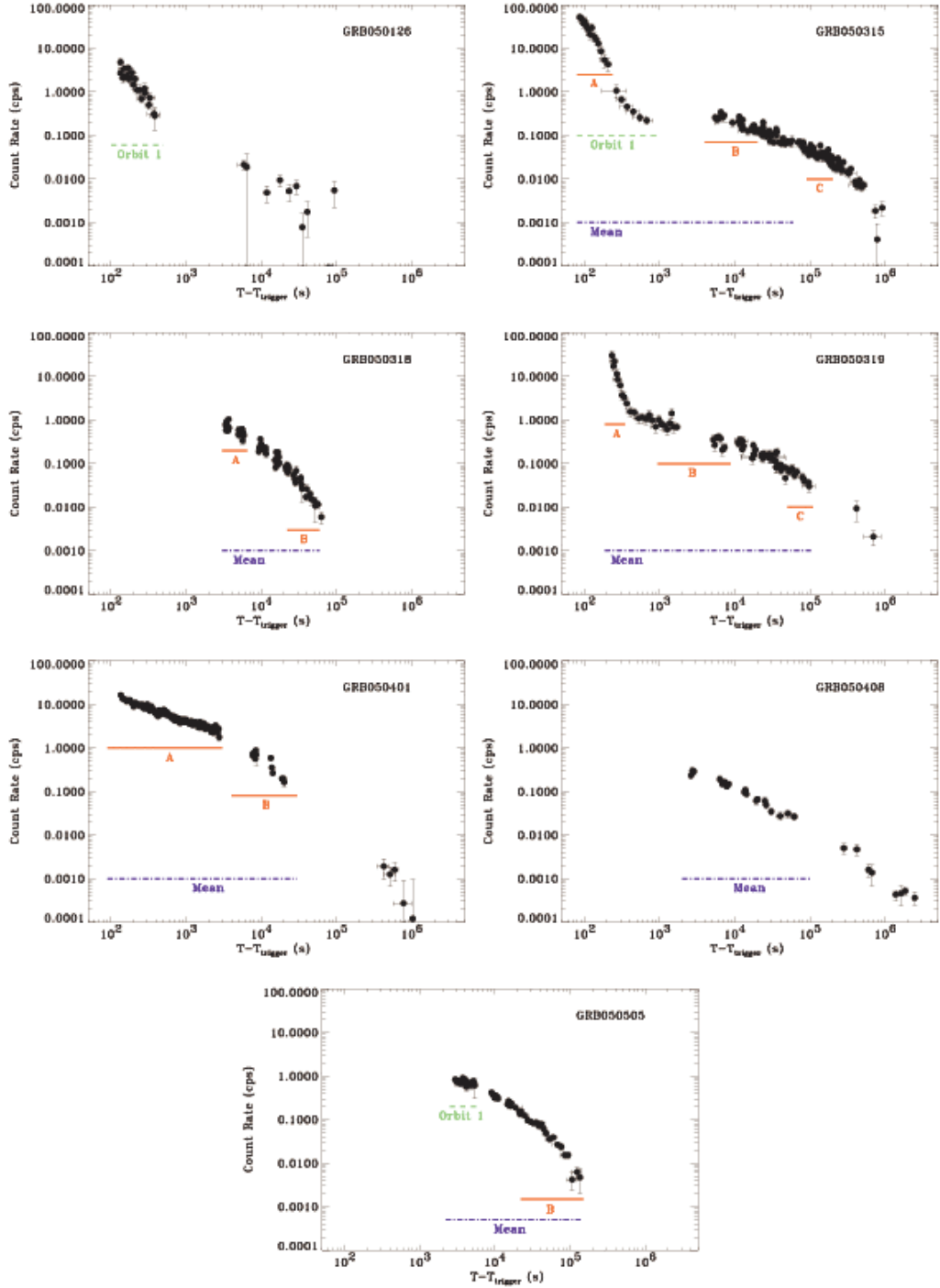

Figure 1. The XRT light curves of the bursts discussed in this paper. For each light curve we have marked the time interval over which the spectrum was measured.

GRB050126
Swift promptly slewed to the burst and XRT started observations 129 s after the BAT trigger. When the burst trigger occurred, XRT was operating in manual state in PC mode. Consequently the first set of data (300 s) is affected by pile-up. To account for pile-up we extracted an annular region of inner radius 3 pixels and outer radius 30 pixels. This region includes 50% of the Encircled Energy Fraction (EEF, Moretti et al. 2004). For more details on the analysis of this burst see Tagliaferri et al. (2005) and Goad et al. (2005).

GRB050315
The observatory executed an automated slew to the BAT position and the XRT began receiving data 80 seconds after the burst. The XRT was in a manual state and performed observations in PC mode only. As in the previous case, the first orbit data are severely piled up. This forced us to consider only data from an annulus with an internal radius of 7 pixels; this provides only 20% of the EEF. For more details on the analysis of this burst see Vaughan et al. (2005).

GRB050318
When BAT detected GRB 050318 the position was within the Swift Earth horizon constraint. Thus, an automatic slew could not be performed until some 55 min after the BAT trigger. After the slew XRT began taking data in PC mode the afterglow was still piled up; therefore we extracted events in an anular region of radii 3 and 25.

GRB050319
For this burst the spacecraft was in automatic mode and executed an immediate slew arriving on target 90 s after the BAT trigger. Due to the brightness of the source, XRT began taking data in WT mode. Because during the first orbit the telescope pointing was 5.7' from the nominal BAT position, the X-ray observations of the GRB fell near the edge of the XRT WT mode frame. This forced us to correct the WT data flux by the fraction of the EEF left out of the window. After the first 50 s XRT switched to PC mode, and up to the end of the 4$^{th}$ orbit data was affected by pile-up. The following observations were taken in PC mode without any pile-up. In order to correct for pile-up we used an annulus of inner and outer radius of 6 and 30 pixels for the first 4 orbits A re-analysis of the prompt light curve by the BAT team found that the GRB did not start at the BAT trigger time reported in GCN 3117, Krimm et al. (2005b), but 137 s earlier. BAT did not trigger at this time as it was slewing from one part of the sky to another. Luckily, a final bright peak occurred which triggered the BAT. In our analysis we adopt the first peak of the prompt emission as the burst trigger time (see Cusumano et al. 2005 for more details).

GRB050401
The spacecraft executed an immediate slew and the first XRT data point was obtained 134.9 seconds after the trigger. XRT was in autostate and began taking data in WT mode. The source was well within the WT window. In subsequent orbits data were collected in PC mode, and the PC source extraction region was an annulus of radii 9 and 30 to account for pile up.

GRB050408
This burst was detected by HETE2. Swift did not slew automatically to the burst position. Instead a ToO (Target of Opportunity) was necessary to re-point the Swift observatory to the burst position. XRT started observing 42 min after the HETE2 trigger. The afterglow had already faded and observations were taken in PC mode.

GRB050505
As for GRB050318, when the BAT triggered on GRB050505, the spacecraft was within the Earth bright limb observing constraint and could not perform an immediate slew: the XRT observations started 47 minutes after the trigger. At this time the source was still piled up and the PC source extraction region was an annulus of radii 3 and 30 to account for pile up.

## 3. Data reduction and analysis

For each burst we extracted the X-ray light curve using the extraction regions described above. In order to arrive at a homogeneous set of light-curves we grouped the counts for each source into bins of at least 30 counts. A typical XRT observation of a GRB lasts between 5 to 8 days with about 10% of the time effectively dedicated to each burst (for the remainder of the time the satellite points either to other GRBs or to other sources). For each orbit (~ 5800 s) a burst is typically observed for about 1000s. This means that for later times, when the source counts are low, we must bin over several orbits to make secure source detections and to satisfy the required number of counts per bin. The light curves for all of the bursts in our sample are shown in Fig. 1. For each burst we take the origin of time as the BAT trigger time (except for GRB050319, see section 3.1) . These light curves show steep decays and slope changes, with count rates spanning 4-5 decades. In order to convert the count-rate light curves to fluxes and then luminosities, we need to calculate the counts to flux conversion factors from the spectral analysis. For each GRB afterglow, the XRT spectra of the source and background were extracted in the same regions used to extract the light curves. We first extracted an average spectrum for each GRB. Ancillary response files were generated with the task **xrtmkarf** within FTOOLS v6.0, to account for the different extraction regions. Spectral redistribution matrices (RMF) were retrieved from the latest Swift Calibration Database) distribution (CALDB 20050405: http://heasarc.gsfc.nasa.gov/docs/heasarc/caldb/swift/).

The spectra were then re-binned with a minimum of 20 counts per energy bin to allow $\chi^2$ fitting within XSPEC 11.3.2. We first performed a fit of the mean spectrum of each burst with an absorbed power law model leaving the column density ($N_H$) as a free parameter. If in the best fit model, $N_H$ was found to be consistent with zero, or below the Galactic value, we froze $N_H$ at the Galactic value. The resultant fits are reported in Table 2. All errors are 90% confidence limits on one interesting parameter.

Table 2

XRT – Spectra 0.2 – 10 keV (1)

| I.D. | Energy Index | $N_H$ ($10^{22}$ cm$^{-2}$) (2) | $\chi^2$/dof | Notes (3) |
|---|---|---|---|---|
| GRB 050126 | 1.26 ± 0.22 | 0.0528 Gal | 1.06 / 8 | Mean Spectrum |
| GRB 050315 | 1.13 ± 0.09 | 0.15 ± 0.02 | 1.58 / 121 | Mean Spectrum |
|  | 1.37 ± 0.17 | 0.13 ± 0.04 | 0.65 / 42 | Orbit 1 |
|  | 1.34 ± 0.15 | 0.13 ± 0.03 | 0.63 / 43 | A |
|  | 0.93 ± 0.15 | 0.15 ± 0.04 | 1.39 / 41 | B |
|  | 0.95 ± 0.13 | 0.15 ± 0.03 | 1.08 / 59 | C |
| GRB 050318 | 0.87 ± 0.09 | 0.04 ± 0.01 | 0.88 / 74 | Mean Spectrum |
|  | 0.99 ± 0.15 | $0.054^{+0.027}_{-0.018}$ | 0.72 / 28 | A |
|  | 1.24 ± 0.28 | $0.076^{+0.038}_{-0.027}$ | 0.99 / 15 | B |
| GRB 050319 | 0.94 ± 0.09 | 0.027 ± 0.009 | 0.90 / 125 | Mean Spectrum |
|  | $1.94^{+0.13}_{-0.20}$ | 0.011 Gal | 281.56 / 526 | A (C-stat) |
|  | 0.79 ± 0.12 | 0.011 Gal | 376.38 / 478 | B (C-stat) |
|  | 0.62 ± 0.18 | 0.011 Gal | 1.33 / 10 | C |
| GRB 050401 | 1.10 ± 0.06 | 0.210 ± 0.02 | 1.11 / 258 | Mean Spectrum |
|  | 1.07 ± 0.06 | 0.210 ± 0.02 | 1.09 / 249 | A |
|  | 1.09 ± 0.06 | 0.21 ± 0.02 | 1.06 / 245 | B |
|  | 1.33 ± 0.40 | 0.29 ± 0.16 | 0.77 / 10 | C |
| GRB 050408 | 1.14 ± 0.19 | 0.25 ± 0.05 | 1.32 / 38 | Mean Spectrum |
| GRB 050505 | 0.95 ± 0.06 | 0.067 ± 0.08 | 1.06 / 174 | Mean Spectrum |
|  | 0.98 ± 0.13 | 0.075 ± 0.022 | 0.89 / 38 | Orbit 1 |
|  | 0.91 ± 0.09 | 0.062 ± 0.013 | 1.11 / 72 | B |

(1) Modeled by a power-law with photoelectric absorption in the 0.2 – 10 keV band
(2) Gal = Milky Way absorption
(3) The interval of time over which the spectrum has been measured is given in Figure 3. 'C-stat' indicates that Cash-statistics were used for model fitting.

For each afterglow we also looked for time-dependent spectral variations, concentrating in particular on identifying spectral changes across breaks in the decay light curves. This was achieved by selecting regions in time either side of the break with sufficient counts to produce statistically meaningful spectra. The selected regions are marked for each spectrum in Figure 1 (correction for pile-up has been applied when necessary). We note that for GRB050319, the spectra extracted in the first two intervals had insufficient counts (due to their short exposure times) for Gaussian statistics to be applied. In this case, we use Cash-statistic (Cash 1979) and fit the un-binned data, instead. As it has been shown by Schmahl (but see also Cash 1979) for counts per bin ≥ 10 the Cash statistic is essentially the same as the $\chi^2$ statistic, while for lower count rates the $\chi^2$ statistics is bad or not reliable.

The values from our fitting procedure, reported in Table 2, indicate that the spectra are consistent within the errors with no spectral variation. Thus, during the evolution of the soft X-ray decay light curve we find no evidence for corresponding spectral evolution. The mean spectral energy index β (where $f_\nu \propto \nu^{-\beta}$), of the observed bursts is <β> = 1.08 ± 0.27. The scatter in β is larger than the typical measurement error implying there exists a physical dispersion of the energy index among bursts, albeit rather small. GRB050319 is an exception to this rule. In this case the spectrum is very soft soon after the burst (indeed it is the softest spectrum measured) and becomes significantly harder at later times. GRB050315 may indicate similar behavior while GRB050318 appears to evolve in the opposite sense. Indeed as stated above there is some physical variation that we may uncover with improved statistics.

## 3.1. BAT light curve analysis.

The BAT data were analyzed with the BAT analysis software FTOOLS v 6.0. For each GRB we also extracted a prompt BAT spectrum to compare it with the XRT spectrum from the steep decline phase. For all bursts the average BAT spectrum is well fitted with a single unabsorbed powerlaw. The fits have been carried out in the energy range 20 – 150 keV where we know, at the time of this writing, the calibration to be the best (see e.g. http://swift.gsfc.nasa.gov/docs/swift/analysis/bat_digest.html). In Table 3 we report the best fit spectral energy index β. Qualitatively there appear to be two types of bursts, one with a harder spectrum (β<0.5), and the other with a softer spectrum (β≥1). In contrast, the mean XRT spectrum immediately following the burst (see Table 2) is always characterized by an energy index β≈1. The indication is that in 3 cases we progress from a rather hard prompt emission to a softer afterglow, but it is not completely clear whether these characteristics are related to the morphology of the light curves. If the missing observations of GRB050318 soon after the prompt emission had been characterized by a rapid decline, then we can safely argue that light curve morphology does play a role. Within the errors the energy index of the BAT spectra for GRB050315, GRB050318, and GRB050319 (Table 3 column 2, Mean = 1.16 ± 0.32) is the same as the energy index of their respective XRT spectra (1.08 ± 0.27). This indicates that a rapid decline of the post prompt emission is characterized by a rather soft spectrum with an energy index that is similar for the hard and soft X rays. GRB050126 is a possible exception.

Table 3: BAT–XRT Energy

| GRB | Energy Index BAT spectra (1) | E $10^{50}$ 15-350 keV erg | E $10^{50}$ XRT Range T1,T2 erg (2) | E $10^{50}$ XRT Range T3,T4 erg (3) | E $10^{50}$ XRT Obs Tot erg (4) | Range (Log sec) of XRT data (5) |
|---|---|---|---|---|---|---|
| 050126 | 0.32 ± 0.18 | 77.3 | 0.59 | 0.44 | 1.25 | 1.77-4.1 |
| 050315 | 1.18 ± 0.11 | 276.7 | 3.88 | 28.3 | 117.4 | 1.46-5.49 |
| 050318 | 1.16 ± 0.13 | 79.4 | … | 4.49 | 4.92 | 3.13-4.41 |
| 050319 | 1.13 ± 0.28 | 512.7 | 21.5 | 58.11 | 160.3 | 1.35-5.21 |
| 050401 | 0.13 ± 0.09 | 2748.9 | 43.5 | 126.6 | 327.7 | 1.54-5.42 |
| 050408 | 0.32 ± 0.18 | … | … | 6.5 | 22.7 | 3.07-5.47 |
| 050505 | 0.27±0.15 | 1400.7 | … | 171.8 | 250.7 | 2.75-4.41 |

(1) The spectral fit was done in the band 20 – 150 keV where we know the calibration to be the best. The Fluence has been measured in the standard 15 – 350 keV band.
(2) T1 = 50 s, T2 = 200 s after the BAT trigger.
(3) T3 = 1300 s, T4 = 12600 s after the BAT trigger.
(4) Total Energy observed by XRT.
(5) Gaps during the observations are present.

## 3.2. XRT light curve analysis

To assess whether there are common features in the X-ray light curves, we converted them from count rates to fluxes, using for each burst the conversion factors derived from the spectral analysis of their average spectra. We then computed the luminosity in an $H_0 = 70$ km/s/Mpc, $\Omega_m=0.3$, $\Omega_\Lambda=0.7$ Universe. The light curves are presented in Figure 2. After correcting the temporal scale to the source rest frame, the XRT observations typically begin ~20 to 30 seconds after the BAT trigger. In Figure 2, the squares mark the average BAT luminosity observed during the prompt phase. For each burst we only report the average prompt luminosity estimated over the $T_{90}$ spectrum to avoid overcrowding.

In order to model the XRT light curves shown in Fig. 2, we employ three different methods. In the first method we fit the light-curve with a power law with either one or two breaks. For the former the analytic expression is:

$$F(t) = \frac{K}{\left(\dfrac{t}{t_{breack}}\right)^{\delta_1} + \left(\dfrac{t}{t_{breack}}\right)^{\delta_2}}$$

where $\delta_1$ and $\delta_2$ represent the two different slopes. For the latter we adopt:

$$F(t) = At^{-\delta_1} + \frac{K}{\left(\dfrac{t}{t_{second\ break}}\right)^{\delta_2} + \left(\dfrac{t}{t_{second\ break}}\right)^{\delta_3}}$$

These equations, however, imply at any time the superposition of two (in the first case) and three (in the second case) signals, so that the true value of the slopes is biased if the slopes evolve with time. In practice this turns out to be a reasonable approach. The results of this method are reported in the first row of Table 4 for each burst. The second method we consider involves estimating the slope of the light curves within intervals located away from the apparent positions of the breaks in the light-curves. We do this by excluding the data close to the break times and fitting the remainder of the light-curves with powerlaws (see Table 4, row 2 for each burst). Finally, we also estimated slopes by deriving the tangent to the fit at various points along the light-curve (values reported in Table 4, row 3 for each burst).

The location of the break-time has the highest uncertainty. Perhaps the most unbiased procedure for estimating light curve parameters is: a) fit the observed light curve with any fitting function (e.g. a polynomial), b) estimate the slope from the first derivative of the fitting function, and c) estimate the break-time by measuring the position of maximum curvature. The curvature is given by the following relation where *f(t)* is the function fitting the observations:

$$K = \frac{\left|\dfrac{\partial^2 f(t)}{\partial t^2}\right|}{\left(1 + \left(\dfrac{\partial f(t)}{\partial t}\right)^2\right)^{3/2}}$$

Although estimating the errors for method 3 is rather cumbersome, it does have the advantage of allowing high temporal resolution.

Table 4 shows that the light curves have the fairly common and well known behavior, consisting of a rather fast decay interrupted by light curve breaks.

During the very early phase there are three GRBs which show a steep decay, whereas the light curve of GRB 050401 (z=2.9) differs considerably. In this case we note that the early temporal slope for this burst is similar to what was observed after the first break in the other three GRBs. Unless we missed an early, fast, and brief decay soon after the BAT prompt phase, this may indicate that we have at least two distinct types of light curves.

Table 4. Slopes (-α) and $t_{break}$ of the light curves.

| | Temporal slope index Phase 1 | Break Log sec First (1) | Temporal slope index Phase 2 | Break Log sec Second (1) | Temporal slope index Phase 3 | Notes |
|---|---|---|---|---|---|---|
| GRB050126 | $-3.40^{+0.94}$ | $2.11^{+1.7}$ | -0.94±0.36 | … | … | |
| | -2.68±0.18 | 2.44±0.83 | -0.66±0.35 | … | … | |
| | 2.67±0.22 | 3.0±0.40 | -0.56±0.05 | … | … | |
| GRB050315 | -4.02±0.27 | 2.04±0.13 | -0.20±0.05 | 4.5±0.28 | -1.65±0.18 | (2) |
| | -3.25±0.19 | 2.16±0.19 | -0.14±0.069 -0.67±0.11 | 3.45±1.26 4.97±1.46 | -2.65±0.39 | |
| | -3.28±0.89 | 2.45±0.48 | 0.14±0.26 -0.67±0.26 | 3.21±0.14 4.77±0.30 | -2.47±0.55 | (3) |
| GRB050318 | … | … | $-0.89^{+0.38}$ | 3.95±0.55 | -2.63±0.50 | |
| | … | … | -1.04±0.09 | 3.89±0.97 | -2.41±0.21 | |
| | … | … | -1.05±0.07 | 3.7±0.05 | -2.38±0.28 | |
| GRB050319 | -7.64±3.83 | 1.89±0.12 | -0.50±0.08 | 4.24±0.39 | -2.07±0.06 | |
| | -6.15±0.27 | 1.94±0.13 | -0.58±0.04 | 3.89±0.7 | -1.23±0.06 | |
| | -6.58±1.1 | 2.16±0.13 | -0.52±0.18 | 3.56±2.02 | -1.18±0.16 | |
| GRB050401 | -0.54±0.05 | 3.35±0.2 | -2.10±0.26 | … | … | |
| | -0.59±0.02 | 3.13±0.31 | -1.63±0.05 | … | … | |
| | -0.64±0.05 | 2.6±0.16 | -1.76±0.40 | … | … | |
| GRB050408 | … | … | -0.56±0.15 | $5.05^{+1.2}_{-0.5}$ | -1.54±0.2 | |
| | … | … | -0.67±0.09 | 4.48±1.15 | -1.35±0.09 | |
| | … | … | -0.66±0.03 | 4.42±0.08 | -1.24±0.19 | |
| GRB050505 | … | … | -0.71±0.18 | 3.84±0.46 | -2.45±0.40 | |
| | … | … | -0.79±0.04 | 3.37±0.72 | -2.29±0.30 | |
| | … | … | -0.82±0.16 | 3.40±0.18 | -2.29±0.30 | |
| Composite Curve | 2.52±0.009 | 2.22±0.31 | -0.69±0.07 | 4.91±0.04 | -1.99±-0.03 | |

(1) Rest frame
(2) This is a very complex light curve. The slope at the beginning increases with time more or less as for the other light curves. After ~200 s it decreases rather slowly and, here is the main anomaly, after ~800 s the light curve becomes flat, decreasing again after about 8000 s (see Figure 1).
(3) This curve presents a short plateau so that for a small time interval we measure a slope of approximately zero

To reiterate, it is whether the initial slope of GRB050401 represents the beginning of the afterglow or the slope we observe after having missed a steeper decline between 4 s and 30 s (rest frame). We disregard for the moment this second possibility as: a) we have no evidence from the data at hand, and b) a missed initial steeper decline would have meant a flux higher by at least a factor of 10, bright enough to be detected by the BAT.

The BAT light curve of GRB050319 is characterized by two energy peaks separated by about 137 seconds, the first peak has an equivalent (referred to the XRT band) mean flux of $5.7 \times 10^{-8}$ erg cm$^{-2}$ s$^{-1}$ compared to a flux of $2.9 \times 10^{-8}$ erg cm$^{-2}$ s$^{-1}$ in the second peak. The formal trigger, and therefore our reference of time, is referred to the start of the first peak of the prompt emission detected by BAT. By using the first peak as our reference time, we find the steepest slope for any of the bursts (see Table 4). Conversely, if we use the beginning of the second peak as the trigger time, i.e. 137 seconds after the start of the first peak, the light curve of GRB 050319 becomes slightly less steep and very similar to the light curve of GRB050318. As we will detail later, we are not aware of any plausible physical explanation for decay slopes as steep as -6 or -7 and this may be giving a clue as to the physical trigger time (by which we mean the real beginning of the afterglow or onset time). This finding highlights the importance of understanding how to observationally measure, or derive, the time at which the afterglow begins.

Indeed, there may exist a bias in our analysis since the value of the early slope is far more strongly dependent upon the choice of $t_o$, than the slope determined at later times. Plotting the XRT data with a time beginning at the time of the BAT trigger, essentially means plotting the function $L = K (t - t_o)^{-\alpha}$ with $t_o = t_{trigger} = 0.0$. Clearly this has no physical meaning for our purposes since the beginning of the XRT light curve, i.e. the beginning of the afterglow in the current fireball model, may not coincide with the BAT trigger, which is an operational time related to the beginning of the prompt emission. For instance, the question that is raised by the examination of the rest frame GRB light curves is whether or not the very beginning of the light curve has a constant decay slope and that the observed differences are partly due to an incorrect choice of $t_o$. Further development of this argument requires detailed simulations of the burst and early afterglow emission and is beyond the scope of the present work.

By way of illustration, consider the case of GRB050319. For this source we had to revise the time of the burst trigger as BAT triggered on the second of two peaks in the BAT light-curve, the first occurring some 137 seconds earlier. If we use the first 100 seconds of the XRT light curve (rest frame) we find a slope of $\alpha = 2.87$ using the first ``erroneous'' trigger as $t_o$ (in this case XRT observations would start 95.2 s after $t_o$). A slope of $\alpha = 5.31$ is measured using the corrected BAT trigger time (in this case the XRT observations would start 232 seconds after $t_o$). If we go to the extreme and put $t_o$ at the end of the BAT prompt emission with an observed $T_{90} = 160$ seconds, we derive a slope $\alpha = 2.43$ (XRT observations would start 72 seconds after $t_o$ in this case). The initial slope changes but the shape of the light curve remains the same. Note that GRB050319 with the largest $T_{90}$ is the worst case. For all the other GRBs of the sample $T_{90} < 100$ seconds (observer frame) so that the possible range of $t_0$ is small and the the slope does not change much.

Due to our current limited understanding of this point, the BAT trigger remains our best choice of reference.

The most striking feature shown by Figure 2 is the similarity in the decay of the luminosity from the BAT to the XRT data once we smooth over the details of individual light curves. Figure 3 indicates the different behavior shown by the light curve of GRB050401 where the afterglow begins with a mild slope $\alpha \approx 0.6$. All of the XRT light curves seem to point toward the mean flux observed by BAT (square symbols in Figure

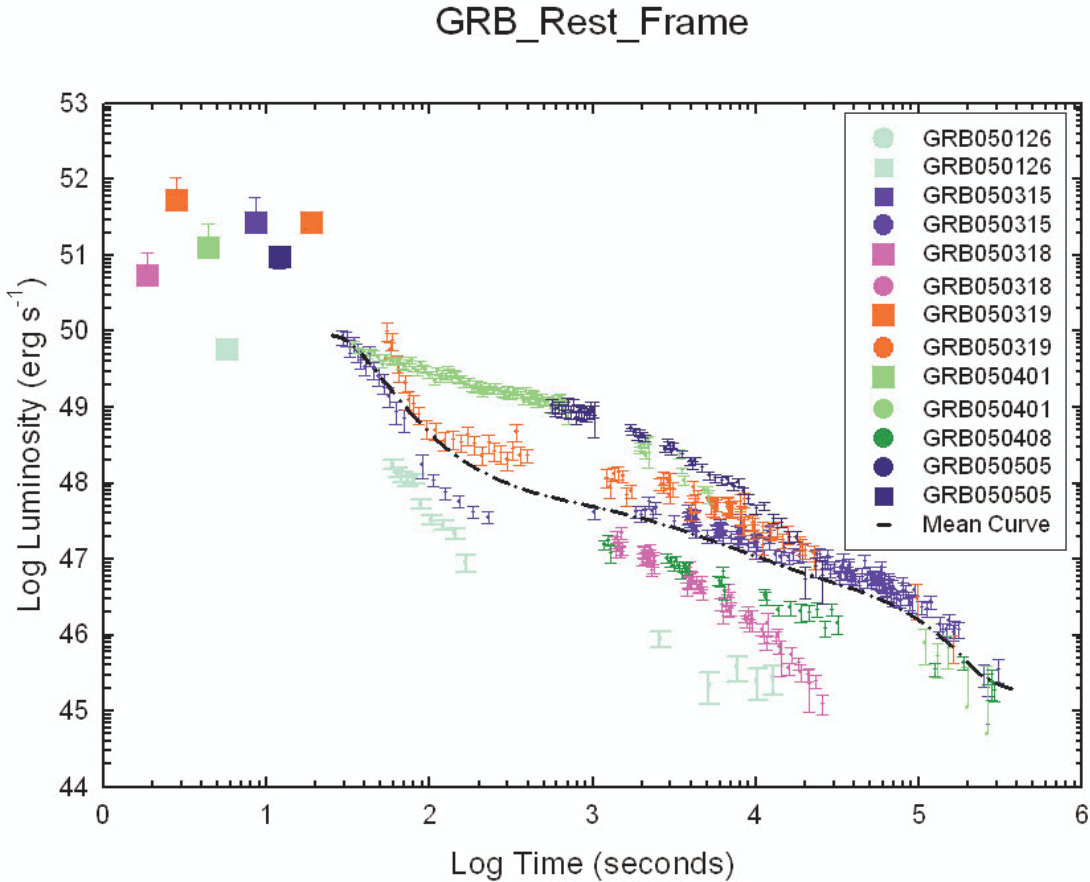

*Figure 2: The light curves have been plotted in the rest frame of each single burst. For the origin of time we used the trigger time as given by the BAT instrument. Squares refer to the mean flux observed by BAT during the burst and converted via the BAT spectrum to the energy band of XRT, circles refers to the observed flux in the band 0.2 – 10 keV. The dot dashed line is the mean curve as described in the text.*

2). This suggests a unique model as a source of the luminosities we observe for light-curves showing an early rapid decay. For this reason and rather than reporting average values for each light curve, we have constructed a composite light curve, formed from summing the observations in the rest frame of each source (Figure 2 - dot – dashed line). This exercise is illustrative only and should be carried out on a larger sample for the different GRB types. In particular, the light curve of GRB050401 was excluded because it displays a rather different set of characteristics over this time interval, being flatter early on before breaking to a steeper slope. The XRT light curve of GRB050505 does not cover the phase soon after the prompt emission. This is suggestive of at least two types of decay behavior. Our composite light curve is designed to be representative of the `type 1' light curves (GRB050315 type). GRB050401 may be the prototype of the`type 2' light curves (see also GRB050128, Campana et al., 2005).

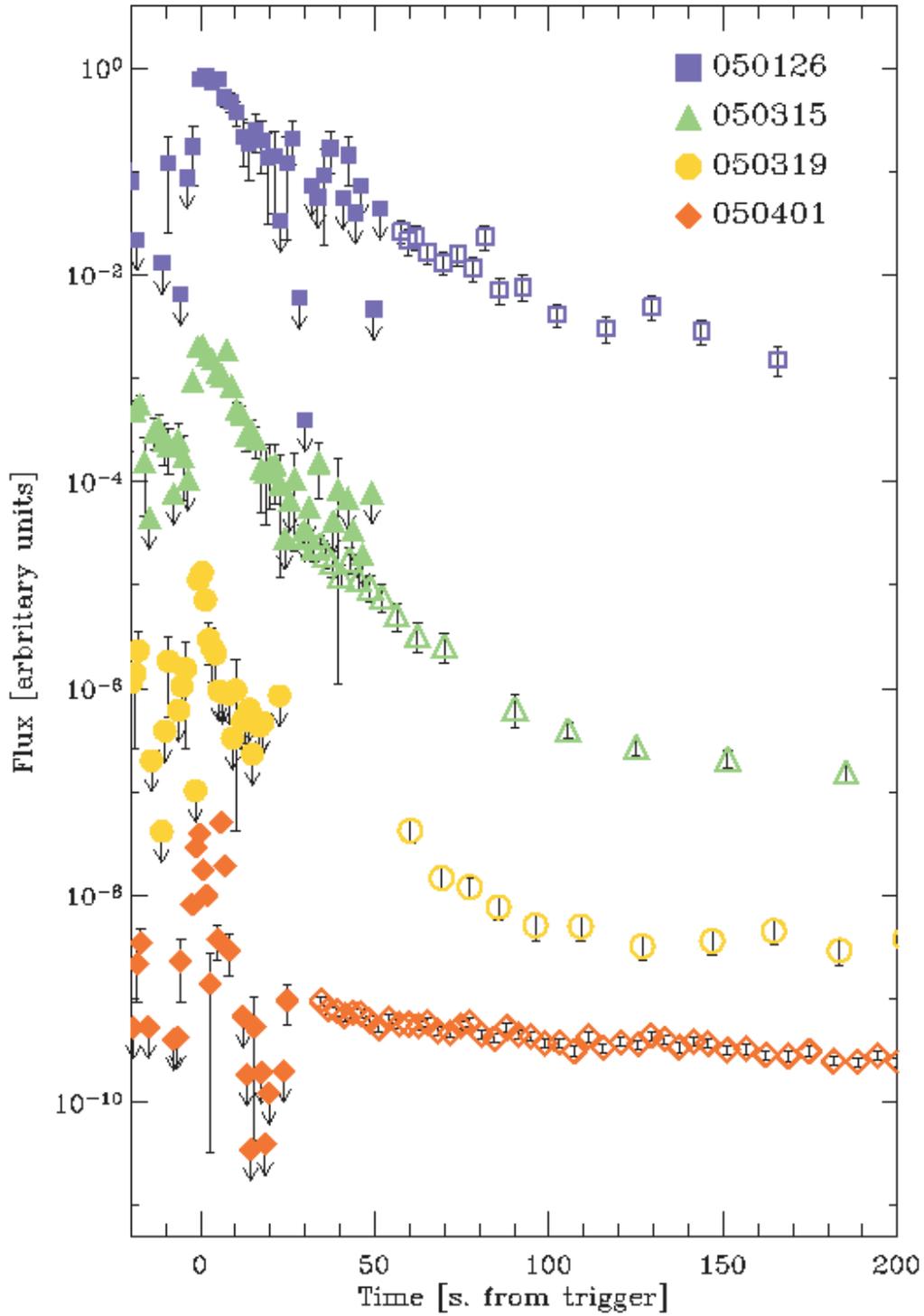

*Figure 3. Light curves for 4 of the 7 GRBs observed by Swift for which we have redshift and early XRT observations. In the case of GRB 050315 the BAT and XRT observations overlap.*

Of some interest is to compare the luminosity of the prompt to the luminosity at the very beginning of the afterglow. For the reference time, we choose 30 s as the best compromise for measuring the afterglow emission without using a large extrapolation. The XRT flux at 30 s (rest frame) after the trigger has been computed extrapolating the XRT observations, but only for those curves observed within 60 seconds (rest frame time). The error in the first point of the XRT light curve ranges from 10% for GRB050401 to 25 % for GRB050319. For the 4 objects with early afterglow observations (GRB050126, GRB050315, GRB050319, and GRB050401) we measure a flux ratio (XRT at 30s) / BAT = 0.16, 0.028, 0.22, 0.06 respectively, with an error, accounting for the errors in the XRT data points given above, smaller than 20%. This may be an indication that the afterglow luminosity correlates with the luminosity of the prompt emission .

To clarify this point, we estimated the energies emitted in soft X-rays (0.2-10 keV) at various intervals to compare with the energy emitted in the hard X-rays (15-350 keV). Most of the GRB light curves were observed in the range 50 – 200 s (rest frame) and 1300 – 12600 s (rest frame). Therefore we selected these time intervals to characterize the light curves and to estimate the energy emitted during these intervals in soft X-rays. In this way, we do not need to extrapolate or derive a mean light curve, and the measurements directly reflect the observations. Integration is in all cases carried out numerically. The results are presented in Table 3. Here we give for each burst (column 1) the spectral index of the prompt emission (column 2) and its fluence (rest frame), column 3. In column 4 and column 5 we report the fluence observed in the XRT band in the time intervals 50 to 200 seconds, and 1300 to 12600 seconds after the BAT trigger. Column 6 gives the XRT fluence over the whole of the observations, and in column 7 we report the start and end time of the observations (Log seconds) in the rest frame of the source.

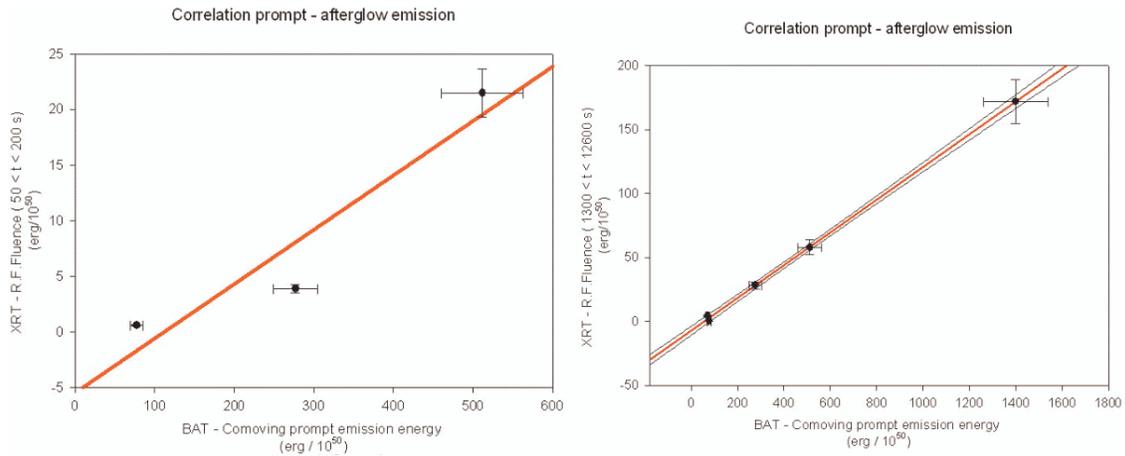

*Figure 4. Correlation between the isotropic energy emitted ( in units of $10^{50}$ erg) during the prompt phase and the isotropic energy emitted in the X-ray band (0.2 – 10 keV) during the afterglow. In the left we show the energy emitted – rest frame fluence - in the time interval 50 s < t < 200 s as a function of the prompt emission (as measured by BAT). On the right we show the same correlation for XRT over the time interval 1300 s < t < 12600 s.*

The correlation between the prompt phase and afterglow for these time intervals is plotted in Figure 4. We do not include GRB050401, as these bursts displays markedly different light curve decay characteristics immediately following the prompt phase. For GRB050505 we do not have data immediately following the burst, apparently the light curve has similar characteristic of GRB050401 and however for this burst the 1300 – 12600 rest frame fluence fits the correlation.

We assigned to each measure in Figure 4 an error of 10%, this is an upper limit. The accuracy of the interpolation of the light curves in the region 50 – 200 s and 1300 – 12600 s respectively and the related integration is much higher, so that a 10% error essentially reflects the uncertainty in the measured fluxes. The probability of obtaining data as different or more different from the fits above is (ANOVA P value) 0.11 for the plot on the left and 0.002 for the plot on the right.

The correlation seems to be very robust albeit for a small sample of bursts. That is the energy in the afterglow is tightly correlated with the energy emitted in the 15 – 350 keV band during the burst.

For those bursts for which we have XRT observations within a few hundred seconds of the prompt emission we can compare the total fluence observed in soft X-rays with the total fluence observed at higher energies, i.e. the ratio of column 6 and 3. We find that the typical "afterglow" X-ray emission ranges form a minimum of 16% (GRB050126) to a maximum of 42% (GRB050315) of the energy emitted during the prompt phase.

## 4. Discussion

The fireball model, Meszaros and Rees Ap. J. 476, 232 (1999), Sari and Piran, Ap. J. 517, 1109 (1999), is in very good agreement with the observations of the single bursts as shown by previous observations of GRBs, and more recently by the Swift satellite. By investigating the early decline phase, immediately following the prompt emission, we can gain information on the energy balance and ultimately an insight into the mechanism of the internal engine and the nature of the progenitor.

As mentioned previously, (Tagliaferri et al. 2005; Burrows et al. 2005b; Cusumano et al. 2005) a very steep initial decline could arise in reverse or internal shocks. After the shock crossing (heating), the shocked material cools radiatively and adiabatically. Radiation from shocked material decays very fast. Once the cooling frequency drops below the observed frequency, the flux decays exponentially. However, the angular time delay effect prevents abrupt disappearance. Off-axis emissions (high latitude emissions) begin to dominate. The expected relation between the temporal decay index and spectral index is universal for synchrotron emission from spherical fireballs (or jets with an opening angle much larger than the relativistic beaming scale), and is given by $\alpha=2+\beta$ (Kumar and Panaitescu 2000) where $F_\nu \propto t^{-\alpha} \nu^{-\beta}$. The typical value of $\beta$ is about 0.5 -1., the maximal decay index could be $\alpha \sim 3$. If the decay is steeper than 3, as may be the case for GRB 050319, we possibly have to argue for a highly collimated jet. Since there is no material at high latitudes, the luminosity would drop faster. In the case of GRB 050319 we may be able, as stated, to decrease the slope by using a $t_0$ that is different from the trigger time. Note, however, that even assuming a $t_0$ coincident with the end of the BAT prompt emission (an extreme that is physically untenable) we have a slope $\alpha = 2.43$.

The second point we make in this work is that, as expected from the preliminary data, we have different types of GRBs that are characterized by a different initial decay. Two types of light curves have been characterized in this sample: `type 1' with a steep early decay followed by a flatter decay preceding another steep decay, and `type 2' which start with a mild early decay (see also the light curve of GRB050128 by Campana, 2005a) that steepen at later times. Secondary bursts (flares) have been often observed superimposed to these light curves (Piro et al 2005, Burrows et al., 2005b).. The light curves with an initial shallow decay may be due to a continuously-fed fireball (Zhang and Meszaros 2001).

The most relevant result is probably the tight, albeit small number statistics, correlation between the prompt energy and the energy emitted by the decaying XRT light curve (Figure 3). Integrating over our mean light curve, which may only be indicative of the most common type 1 bursts, and which spans a range in time from 30 seconds to about 73 hours after the burst, we derive an isotropic emission of about $9.0 \cdot 10^{51}$ erg. In the interval for which we computed the correlation (50 – 200 s and 1300 – 12600 s), we have an emission of $7.8 \cdot 10^{50}$ erg and $2.0 \cdot 10^{51}$ erg, respectively.

The internal – external shocks model requires a comparable emission of energy during the prompt and the afterglow phases. In contrast, the observations show that the energy in the afterglow is at most 40% of the energy we observe during the prompt phase. Clearly, we require better statistics in order to refine the details of the model and ISM. Indeed, here we are referring only to the energy emitted in the X-ray band. We certainly have a broader range of energy related to the prompt phase and it seems likely that the energy transferred to the ISM is more than what we observe. For a theoretical discussion on the relationship, if any, between the prompt and afterglow energy see Lloyd-Ronning & Zhang, (2004).

The sharp decrease of the afterglow seems to be correlated with the morphology of the prompt emission as observed by the BAT instrument. The prompt emission is in all cases (GRB050126, GRB050315, GRB050319) characterized by a sharp rise of the light curve and smooth decay. In stark contrast for GRB 050401 we have two nearly symmetrical peaks, with large fluctuations. The BAT light curve of GRB050319 also shows an earlier (137 seconds) peak that does not show a fast rise. In this case the $t_0$ of the XRT light curve may be related to the second peak. GRB050318 and GRB050505 were not observed by XRT immediately after the trigger, while for GRB050408 we do not have useful BAT information. The indication therefore is that, and this will be clarified with an analysis of a larger sample, the steep decay of the X-ray curve may correlate with one or other of the prompt peak phases.

The XRT spectra are the best compromise between the temporal resolution of the break timescale and the photon statistics needed to measure spectral parameters. In this small sample we do not have evidence of varying $N_H$; this issue will be treated separately in a much larger sample. The main result of this analysis is that the observed energy index shows little, if any, variation among the bursts. For our sample we derive a mean energy index $=1.10 \pm 0.29$, where the error on an individual derivation, as listed in Table 2, range between 0.06 and 0.4. In the case of GRB050319, however, there is a significant variation of the energy index after the first break. The spectrum evolves from very soft (3 $\sigma$ above the mean value) to hard (about 1 $\sigma$ below the mean value), which may be due to the new injection of energy from the second burst.

## Acknowledgments

We are grateful to Dino Fugazza who helped us in preparing the manuscript and to Pawan Kumar for insightful discussions. The work is supported in Italy by funding from ASI on contract number I/R/039/04, at Penn State by NASA contract NAS5-00136 and at the University of Leicester by PPARC contract number PPA/G/S/00524 and PPA/Z/S/2003/00507. We acknowledge in particular all those member of the Swift Team at large who made this mission possible. This goes from the building of the hardware, the writing of the software, the operation at the Mission Operation Centre and the performance of the ASI ground segment at Malindi, Kenya.